\documentclass[aps,preprintnumbers,preprint,superscriptaddress,nofootinbib,tightenlines]{revtex4}
\pdfoutput=1 
\usepackage{amsmath,latexsym,amssymb,graphicx,hyperref,color,enumerate,url}

\usepackage{slashed}

\hypersetup{colorlinks,citecolor= nicegreen,linkcolor= nicered}
\usepackage{color}
\definecolor{nicered}{rgb}{0.7,0.1,0.1}
\definecolor{nicegreen}{rgb}{0.1,0.5,0.1}

\def\ben{\begin{enumerate}} 
\def\een{\end{enumerate}} 

\def\beq{\begin{equation}}
\def\eeq{\end{equation}}
\def\bea{\begin{eqnarray}}
\def\eea{\end{eqnarray}}

\begin{document}

\title{On the Higgs Fit and Electroweak Phase Transition}

\author{Weicong Huang}
\email{huangwc@itp.ac.cn}
\affiliation{State Key Laboratory of Theoretical Physics and  Kavli Institute for Theoretical Physics China (KITPC),  Institute of Theoretical Physics,  Chinese Academy of Sciences, Beijing 100190, P.\,R.\,China}
\author{Jing Shu}
\email{jshu@itp.ac.cn}
\affiliation{State Key Laboratory of Theoretical Physics and  Kavli Institute for Theoretical Physics China (KITPC),  Institute of Theoretical Physics,  Chinese Academy of Sciences, Beijing 100190, P.\,R.\,China}
\author{Yue Zhang}
\email{yuezhang@ictp.it}
\affiliation{International Center for Theoretical Physics, Trieste 34014, Italy}

\date{\today}

\begin{abstract}
We consider the Higgs portal through which light scalars contribute both to the Higgs production and decay and Higgs effective potential at finite temperature via quantum loops. The positive Higgs portal coupling required by a strongly first order electroweak phase transition is disfavored by the current Higgs data if we consider one such scalar. We observe that by introducing a second scalar with negative Higgs portal coupling, one can not only improve the Higgs fits, but also enhance the strength of first order EWPT. We apply this mechanism to the light stop scenario for electroweak baryogenesis in the MSSM and find a light sbottom could play the role as the second scalar, which allows the stop to be relatively heavier. 
Non-decoupled effects on the Higgs or sbottom self-interactions from physics beyond MSSM are found to be indispensable for this scenario to work. A clear prediction from the picture is the existence of a light sbottom (below 200\,GeV) and a light stop (can be as heavy as 140\,GeV), which can be directly tested in the near future.

\end{abstract}



\maketitle

\section{Introduction}
This year has witnessed the announcement of the Higgs boson discovery by the ATLAS~\cite{ATLAS} and CMS~\cite{CMS} collaborations. 
A crucial test of the electroweak symmetry breaking (EWSB) mechanism
is to precisely measure the Higgs couplings to the other Standard Model (SM) particles. A SM-like Higgs measurement can put indirect constraints on new physics~\cite{Batell:2011pz,Carmi:2012yp, Azatov:2012bz, Espinosa:2012ir, Giardino:2012ww, Low:2012rj, Giardino:2012dp, Corbett:2012dm, Espinosa:2012im, Carmi:2012in,Plehn:2012iz} close to the electroweak scale if they couple to the Higgs field. Once those constrains have been obtained, we shall start to explore opening questions related to EWSB.

One big question falls into the above category is the fate of electroweak baryogenesis (EWBG)~\cite{Kuzmin:1985mm}. 
In this scenario, baryon asymmetry is generated at the electroweak scale through sphaleron transitions  
where a strongly first order electroweak phase transition (EWPT) is required to
create the departure from equilibrium and prevent the wash out of baryon asymmetry. 
Therefore, the nature of EWPT would serve as the first window to test the viability of EWBG.  
The SM is known to fail to provide the above condition and new physics must be introduced through the Higgs portal.
Since our current data from LHC on the Higgs production and decay already suggest the preferred parameter space of Higgs couplings from their global fits, it is natural to map the nature of EWSB at the zero temperature to the one at the high temperature, and seek for the impact on EWBG.

\begin{figure}[t]
\centering
\includegraphics[width=0.8\textwidth]{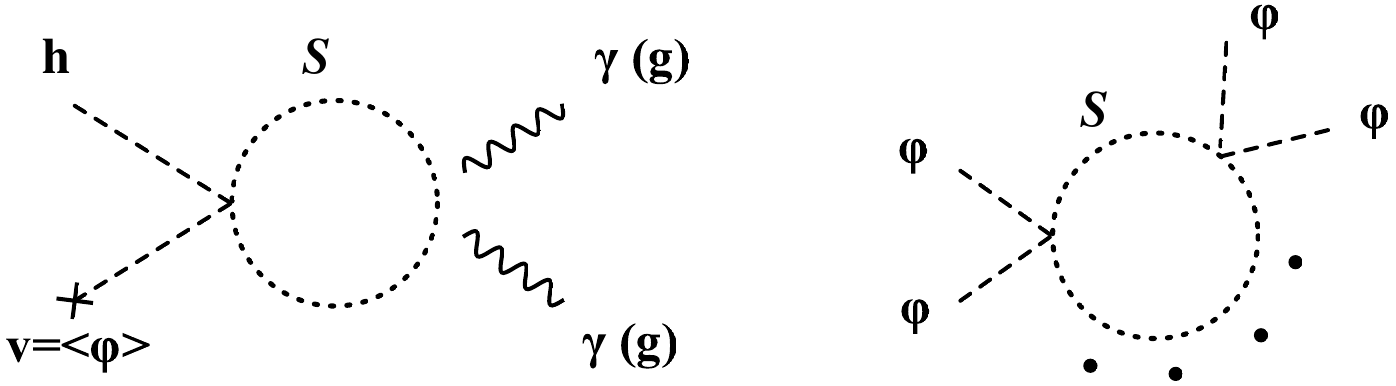}
\caption{Left panal: Contribution of exotic scalar $S$ to the Higgs coupling to photons and gluons. 
Right panel: thermal contribution from $S$ loop to the Higgs potential at high temperature.}\label{1}
\end{figure}

Generally, new physics that connect LHC Higgs signal to EWPT can manifest in several ways. 
New scalars could contribute to both processes through the virtual loop effects (see Fig.\;\ref{1}), or via the mixings with the Higgs boson~\cite{comment1}. 
The simplest example which triggers a strongly first order EWPT is to add a scalar field $S$ coupling to the Higgs. Consider the general scalar mass term
\bea\label{highTmass}
m_s^2(\phi, T) = m^2 + \Pi_s(T) + \alpha \phi^2 \ ,
\eea
where $\Pi_s(T)$ is the thermal self-energy correction to $S$. 
The thermal contributions to the Higgs effective potential $V(\phi, T)$ include a negative mass cubic term $-{T m_s^3(\phi, T)}/{12\pi}$. At the critical temperature $T_c$ when $\alpha > 0$, this term is the only dominant source which decreases with $\phi$. Its competition with the other $\phi$-increasing terms develops a second degenerate nonzero vacuum for $\phi=v_c$. In fact, this lies at the heart of the light stop scenario for EWBG in the MSSM~\cite{Carena:1996wj}.

Meanwhile, if $S$ is a colorful or electric charged particle, the same portal will also modify the effective $hgg$ and $h \gamma \gamma$ couplings and affect the global fit of the Higgs data. The connection is shown in Fig.~\ref{1}.
For the 125\,GeV Higgs boson discovered, LHC
has seen fewer events in $h\to b\bar b, \tau^+\tau^-$ and some excess in $h\to\gamma\gamma$ channel than SM predictions.
Improving the Higgs global fit over the SM favors a suppressed production and enhanced di-photon branching ratio. This requires $\alpha<0$ for $S$~\cite{comment2}, which indicates a tension between EWPT and LHC Higgs signal~\cite{LTWang}.

The purpose of this work is to examine this tension and seek for possible solutions. Indeed, we can relieve this tension in a generic framework with two scalars $S_1$ and $S_2$. The first one $S_1$ has $\alpha_1 >0$ and facilitates a first order EWPT through its thermal corrections to $V(\phi, T)$. The second one $S_2$ has $\alpha_2 < 0$, which not only improves the global fit of the current Higgs data through its destructive contribution with $S_1$ and the top loop, but also {\it further enhances the strength of EWPT} (characterized by $v_c/T_c$). We define a quantity $F[m_s]$ which measures the contribution to $v_c/T_c$ from different scalar fields and show this effect exists in a large parameter space of $\alpha_{1,2}$ and $m_{s_{1,2}}$ (see Fig.~\ref{toymodel}). Consequently, we expect this mechanism could be widely applied in models of this kind, which include the most popular example, Minimal Supersymmetric Standard Model (MSSM). 

This paper is organized as follows. In the next section, we survey the current status of the global fit to the Higgs data, 
and the impact of multiple new scalars through the Higgs portal. 
In section III, we study the role of the scalars in EWPT, and propose a framework where the tension is relieved between the Higgs fit and strong EWPT, with
two scalars having similar mass but opposite Higgs couplings. 
In section IV, we try to implement this framework in the supersymmetric SM, and realize the constraints within the minimal model.
We discuss possible solutions by extending the MSSM and emphasize the testability of this scenario at the LHC.

\section{Higgs Fit with New Colored States}\label{FIT}

In this section, we discuss the impact on the fit to Higgs data of new states that couple to the Higgs boson.
We start by considering a generic tree-level potential
\begin{eqnarray}\label{V0}
V(H, S) &=& - \mu^2 H^\dagger H +\lambda (H^\dagger H)^2 + m^2 S^\dagger S +\kappa (S^\dagger S)^2 + 2 \alpha (H^\dagger H) (S^\dagger S) \ ,
\end{eqnarray}
where $S$ is a complex scalar, carrying electric charge $Q_s$ and number of colors $N(r_s)$, 
and $\alpha$ characterizes the interaction between the $S$ and the SM Higgs doublet $H$ (with neutral component $\phi$).
At zero temperature, after electroweak symmetry breaking, $\langle \phi \rangle=v=246\,$GeV, the $S$ mass is
\beq\label{mass}
m_s^2 = m^2 + \alpha v^2 \ .
\eeq
The interaction between $S$ and the Higgs boson $h$ is
\beq
\mathcal{L}_{\rm int} = -2 \alpha v h S^\dag S \ .
\eeq
In calculating the production and decay of the Higgs boson, we have to include the new contribution from this exotic state $S$ loop. 
This modifies the corresponding rates, which can be parametrized as~\cite{Carmi:2012in}
\begin{eqnarray}
\frac{\sigma(gg\to h)}{\sigma(gg\to h)_{\rm SM}} = \frac{\Gamma(h\to gg)}{\Gamma(h\to gg)_{\rm SM}} = \frac{\hat c_{g, \rm SM} + \delta c_g}{\hat c_{g, \rm SM} },   \ \ \ \ \ 
\frac{\Gamma(h\to \gamma\gamma)}{\Gamma(h\to \gamma\gamma)_{\rm SM}} = \frac{\hat c_{\gamma, \rm SM} + \delta c_\gamma}{\hat c_{\gamma, \rm SM} } \ .
\end{eqnarray}
The SM coefficients, from $t$, $W^\pm$, $b$ loops, are
$\hat c_{g, \rm SM}= 0.97$, $\hat c_{\gamma, \rm SM}=-0.81$, repsectively~\cite{Carmi:2012in}.
The new physics correction due to the $S$ loops are
\bea
\delta c_g = \frac{C (r_s) }{2} \frac{\alpha v^2}{m_s^2} A_s (\tau_s) \ , \ \ \ \ \
\delta c_\gamma = \frac{N(r_s) Q_s^2}{24} \frac{\alpha v^2}{m_s^2} A_s (\tau_s) \ ,
\eea
where $\tau_i = m_h^2 / 4 m_i^2$, and $A_s (\tau) = 3 [ f(\tau) \tau^{-2} - \tau^{-1}]$,
\bea
f(\tau) = \left\{
\begin{array}{lr} 
11 , & \tau \leq 1 \\
22, & \tau\geq1
\end{array} \right.
\eea
$C (r)$ is the quadratic Casimir of the color representation, Tr$(T^a T^b) \equiv C (r) \delta^{ab}$. 
Notice we have the relations $\delta c_g = \left(9 / 2,\ 2 ,\ 0 \right) \delta c_\gamma/Q_s^2$, for color representations {\bf 8, 3, 1}, respectively. 

Clearly, in order to enhance the $h\to \gamma\gamma$ decay rate or to suppress the $gg\to h$ production, we need $\alpha<0$~\cite{Cohen:2012wg}.

\begin{figure}[t]
\centerline{\includegraphics[width=0.8\textwidth]{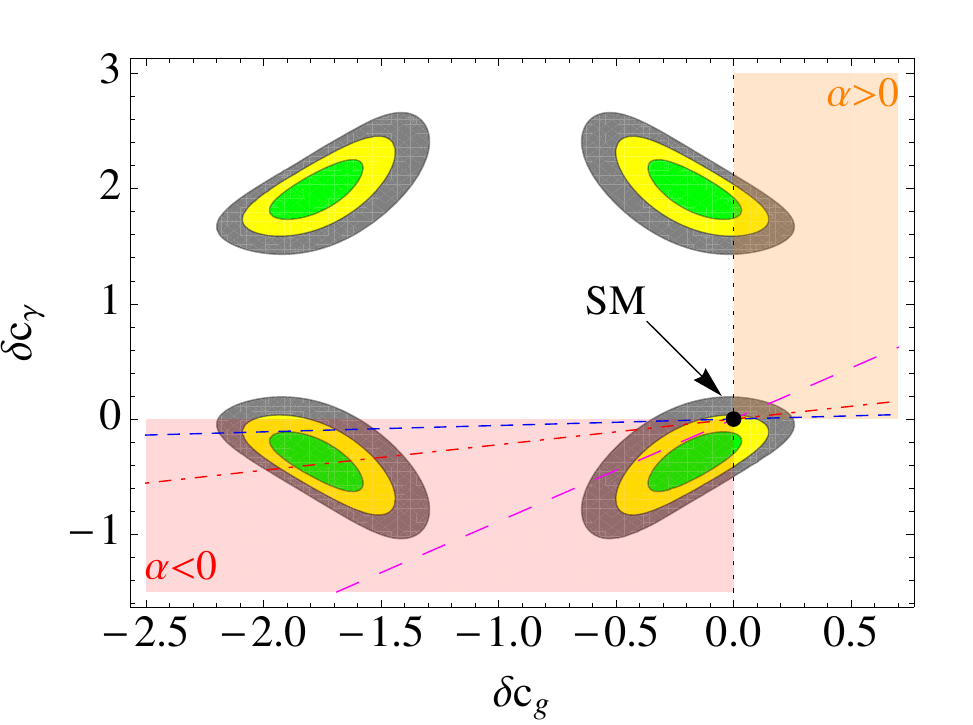}}
\caption{The fit to the LHC Higgs signal in a general two parameter $(\delta c_g, \delta c_\gamma)$ model. The $1, 2, 3 \sigma$ region are show as the green, yellow and gray regions. The colored lines represent the trajectory when a new colored state near the electroweak scale is introduced. Red/Blue/Magenta line: color triplet with $Q_s=2/3, -1/3, -4/3$ respectively. Black (vertical) line: color singlet with $Q_s=-1$.}\label{cgcga}
\end{figure}

In order to quantify the fit to the Higgs signals, we define the $\chi^2$,
\bea
\chi^2 = \sum_{\rm channels} \frac{(\mu_i - \hat \mu_i)^2}{\sigma_i^2} \ ,
\eea
where $\mu_i$ is the signal rate calculated in new physics models normalized to the SM one~\cite{Azatov:2012bz} ($\hat \mu_i \pm \sigma_i$ are the central value and uncertainty measured by experiments)
\bea
\mu_i = \frac{\sum_p \sigma_p(c_g, c_\gamma) \zeta^i}{\sum_p \sigma_{p, \rm SM} \zeta^i} \times \frac{{\rm Br}(h\to i)(c_g, c_\gamma)}{{\rm Br}(h\to i)_{\rm SM}} \ ,
\eea
where $\sigma_p$ is the production cross section of the channel $p$, $\zeta_i$ is the cut efficiency for a particular final state $i$.
In the fit, we take into account of the data released recently in Refs.~\cite{ATLAS, CMS}. 
Throughout the discussion, we assume the Higgs boson has no invisible decay channels. For exceptions, see \cite{Carena:2012np}.

In Fig.~\ref{cgcga}, we show the regions that can fit the Higgs signal in a general two parameter $(\delta c_g, \delta c_\gamma)$ model.
There are four degenerate minimum of $\chi^2$. The SM fit lies between $1\sigma$ and $2\sigma$, and can be improved 
when there are both negative contributions to $\delta c_g$ and $\delta c_\gamma$.
We also show the trajectories that new physics models can cover by adding a single particle with particular color and electric charge quantum numbers, with varying mass and Higgs coupling $\alpha$.
With a light charged scalar (along the dotted vertical line, in black), like the stau, a negative and sizable $\alpha$ clearly profits the fit.

\begin{figure}[t]
\centerline{\includegraphics[width=0.8\textwidth]{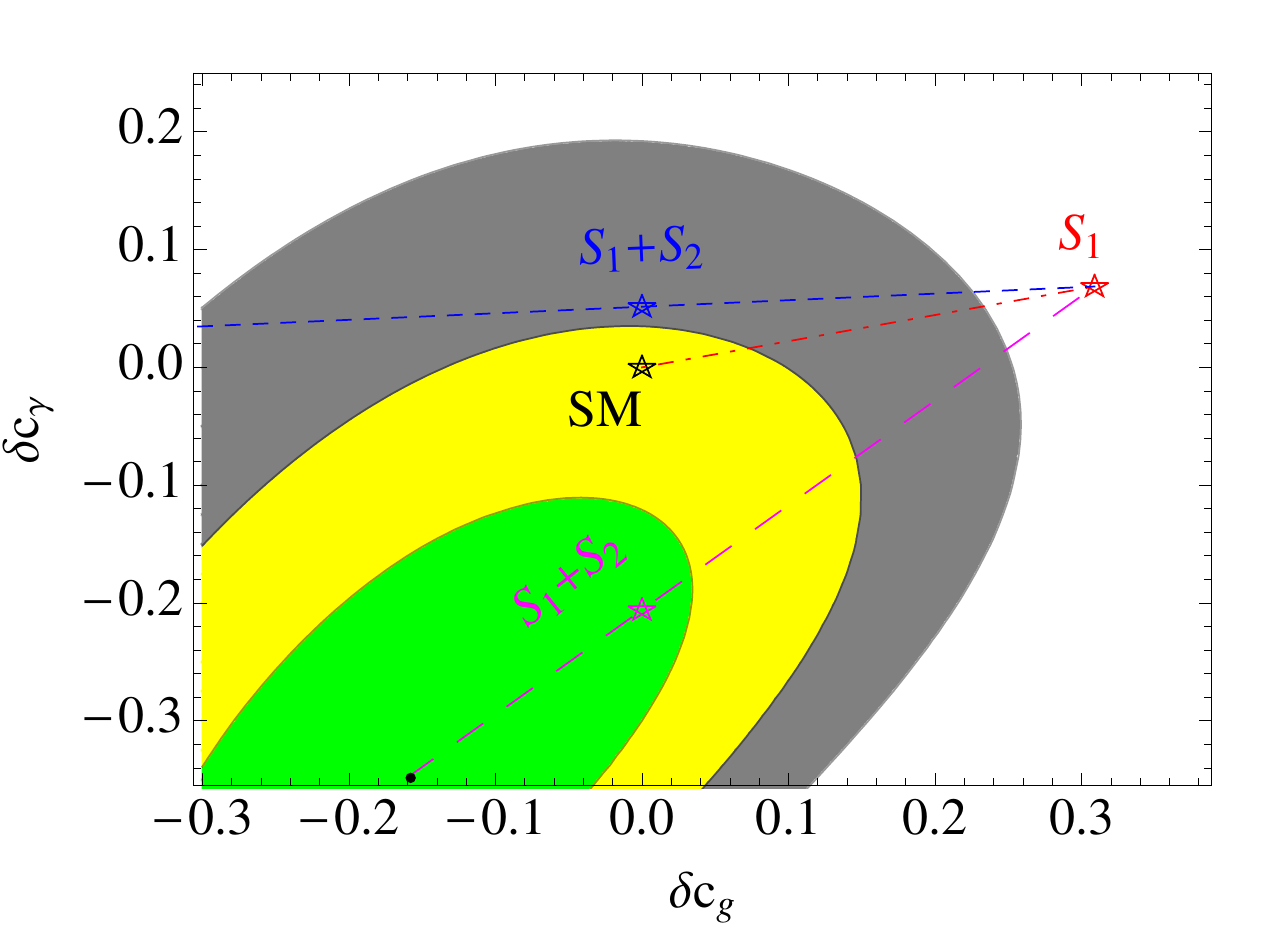}}
\caption{Black star: SM fit. Red star: the fit with one color triplet scalar with electric $2/3$ (stop-like), $m_{1}=120\,$GeV and $\alpha_1=0.5$. 
Blue/Magenta star: adding another color triplet with charge $-1/3$ (sbottom-like) or $-4/3$ (exotica), and mass $m_{2}=120\,$GeV and $\alpha_{2}=-0.5$. 
We find the $Q_{s_2}=-4/3$ state makes the fit even beter than the $-1/3$ one. 
This is because it can more than overcome $S_1$ in $\delta c_\gamma$ and make additional negative contribution.
Here we have assumed no Higgs invisible decay channels open.
}\label{improve}
\end{figure}

Such global fit can be used to constrain new physics scenarios that enter through the Higgs portal. There have been recent analysis in~\cite{Cohen:2012zza, Curtin:2012aa, Carena:2012xa} showing the measured Higgs signal at LHC 
brings severe tension to the light stop scenario for EWBG in the MSSM.
The main obstacle is the top squark (stop)
contributes to the Higgs effective couplings $\delta c_g$ and $\delta c_\gamma$ constructively with the top quark, as they do in the QCD/QED beta functions~\cite{Gillioz:2012se}.
A very light top squark (stop)~\cite{Carena:2008vj}, as required by a strongly first order EWPT,
largely enhances $\delta c_g$ toward an experimentally disfavored direction.
This is the tension mentioned in the introduction.
In Fig.~\ref{improve} we quantify the current status of this tension, in the presence of a single stop-like scalar $S_1$, 
with $\alpha_1\approx0.5$ and mass 120\,GeV. Clearly, the fit (red star) is well outside the $3\sigma$ region. 

This tension can be relieved in two ways. 
First, if $S_1$ is allowed to be heavier without weakening the strength of EWPT, the red star will move toward the SM one (in black), and the fit is improved.
In fact, we find it crosses the $3\sigma$ contour when $m_{s_1}$ is relaxed to around 130\,GeV.

Second, an additional colored scalar $S_2$ may coexist with $S_1$ near the electroweak scale.
If $S_2$ lies in the same color representation as $S_1$, and satisfies $m_{s_2}\approx m_{s_1}$ and $\alpha_2\approx-\alpha_1<0$,
its contribution to the $\delta c_g$ will cancel completely that of $S_1$ and the fit can be significantly improved, as shown in Fig.~\ref{improve}. 

In the next section, we will study the impact on EWPT in the presence of an $S_2$ with $\alpha_2<0$. 
We show that it can assist $S_1$ (with $\alpha_1>0$) and further enhance the phase transition, 
thus allow the latter to be substantially heavier (the star points in Fig. \ref{improve}) . In other words, the same $S_2$ can help to achieve both of the above solutions.
This serves as the main point of this work.

It is natural to expect multiple light color/charged states in new physics models. 
For example, in supersymmetric theories, there are all the scalar partners of quarks and leptons.
When some are light, there could be rich parameter space for them to collaborate and improve the harmony between Higgs fit and EWPT.

\section{Enhancing the Strength of Phase Transition}

As discussed in the introduction, the same coupling $\alpha$ defined in Eq.~(\ref{V0}), 
controls the interaction of scalar $S$ with the Higgs boson $h$, or the high temperature field $\phi$. 
Therefore, it could not only modify the Higgs boson production and decay properties, 
but also contribute to the finite temperature Higgs potential (Fig.~\ref{1}), 
thus play a role in the EWPT.

In order to understand the impact of $S$ in the phase transition, we consider the leading terms in the thermal Higgs potential
\begin{eqnarray}\label{pot}
V(\phi, T) &\approx& \frac{1}{4} \lambda \phi^4 + \frac{1}{2} \left[ -\mu ^2+ \Pi_h(T) \right] \phi^2 - T \left[ E_{\rm SM} \phi^3 + 2 N(r_s) \frac{m^3_{s}(\phi, T)}{12\pi} \right]  \ .
\end{eqnarray}
At high temperature, the masses receive thermal corrections 
\bea
m_h^2(\phi, T)  &=& -\mu ^2 + 3 \lambda \phi^2+ \Pi_h(T)\ , \nonumber \\
m_s^2(\phi, T)  &=& m^2 + \alpha \phi^2 + \Pi_s (T) \ ,
\eea
where the thermal mass corrections are
$\Pi_{h} \approx \left(6\lambda + 2 N(r_s) \alpha +  (9g^2 + 3g'^2)/4 + 3y_t^2\right) T^2/12$,
$\Pi_{s} \approx \left( (2 N(r_s)+2) \kappa + 4\alpha + 4 C_2(r_s) g_3^2\right) T^2/12$, 
with the second quadratic Casimir defined as $T^a T^a = C_2(r) \bf{1}$, and satisfies $N(r) C_2(r) = 8 C(r)$.
For the moment, we neglect the radiative corrections to the effective potential, {\it \`a la} Coleman-Weinberg, and high order terms in $m/T$ for the simplicity of illustration, which will be included later in the numerical studies. 
The thermal mass cubic potential term arises from the daisy resummation and the zero modes of the bosonic fields in the loop~\cite{Quiros:1999jp}.
The critical temperature $T_c$ is defined when two local minima become degenerate. One is the symmetric phase and the other is with non-vanishing $\phi$. The height of the barrier between the two vacua is determined by the $\phi$ cubic term. 

We focus first on the case $\alpha>0$, and solve the degenerate minima conditions, $V(0, T)=V(\phi, T)$ and $V'(\phi, T)=0$. This leads to
\begin{eqnarray}
\frac{N(r_s)}{6\pi}\,T_c \left[ m_s^3(v_c, T_c) - m_s^3(0, T_c) \rule{0mm}{4mm}\right] + \frac{1}{4} \lambda v_c^4= \frac{1}{2} T_c E_{\rm SM} v_c^3 + T_c \frac{N(r_s)}{12\pi} \frac{\partial m_s^3(v_c, T_c)}{\partial v_c} v_c \ ,
\end{eqnarray}
where $v_c=\langle \phi \rangle$ is the vacuum expectation value (VEV) at the critical point.

Therefore, the strength of EWPT or $v_c/T_c$ depends on the SM part $E_{\rm SM}$ and the quantity
\begin{eqnarray}\label{compare}
F[m_s] \equiv \frac{\partial m_s^3(v_c, T_c)}{\partial v_c} v_c - 2 \left[ m_s^3(v_c, T_c) - m_s^3(0, T_c) \rule{0mm}{4mm}\right].
\end{eqnarray}
$E_{\rm SM}$ arises mainly from SM gauge boson contributions and is known to be too small. 
The presence of new scalars can enlarge this coefficient if its coupling to the Higgs field $\alpha$ is sufficiently large. 
We make the requirement $v_c/T_c\gtrsim 0.9$, in order to sufficiently suppress the sphaleron rate in the broken phase, so the following condition should be satisfied
\bea\label{crit}
\frac{{\rm Tr} \left[ N(r_s) F[m_s] \rule{0mm}{3mm}\right]}{v_c^3} \gtrsim 1.2 \left( \frac{m_h}{125\,{\rm GeV}} \right)^2 \ ,
\eea 
where the $\textrm{Tr}$ is a sum over all the diagonal elements of the scalar particle mass matrix. For $\alpha>0$, the thermal potential term proportional to $- T m_s^3(\phi, T)$ decreases as $\phi$ grows, therefore can balance the other positive terms, and
facilitate the development of degenerate vacua of the $\phi$ field around the critical temperature. In order for this effect to be significant, the thermal mass square should be dominated by the $\alpha v_c^2$ term. See \cite{Carena:1996wj, Chowdhury:2011ga} for the explicit examples.

\begin{figure}[t]
\centering
\includegraphics[width=0.8\textwidth]{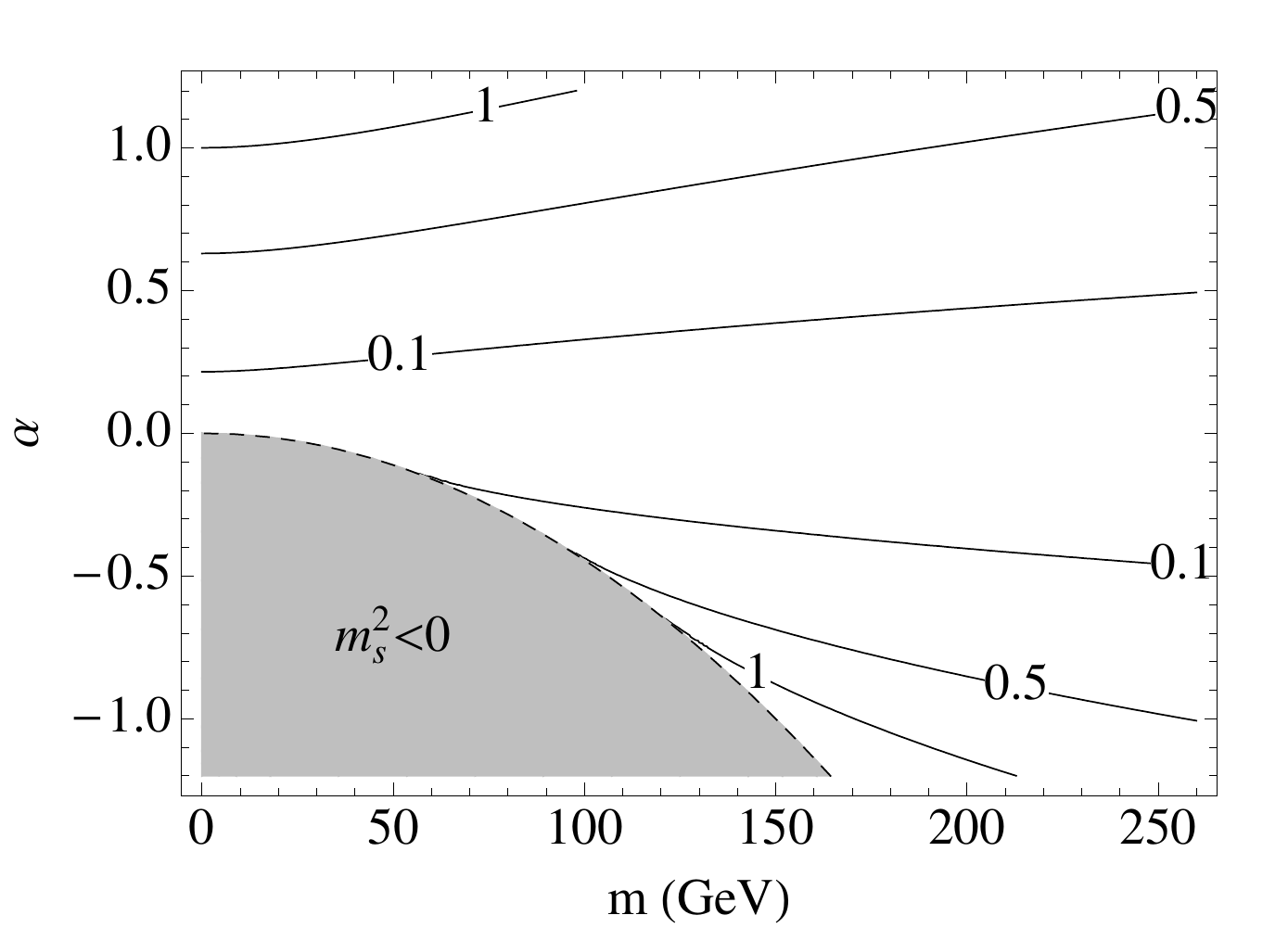}
\caption{$F[m_s]$ in units of $v_c^3$, as a function of parameters $m=\sqrt{m_s^2 - \alpha v^2 + \epsilon_s T^2}$. 
Here the sample value of $v_c$ is taken to be $150\,$GeV. One can see $F[m_s] /v_c^3 \sim \mathcal{O}(1)$ can be achieved for either positive $\alpha$ and small $m^2$, or negative $\alpha$ with low values of $m_s^2(T)$.}\label{positive}
\end{figure}

We point out, interestingly, $F[m_s]$ is always positive, for
\begin{eqnarray}\label{9}
m_s^2(v_c, T_c) = (m^2  + \Pi_s (T_c) ) + \alpha v_c^2 \ ,
\end{eqnarray}
and $\alpha$ taking arbitrary sign. For this quantity to be large enough, as demanded by strong phase transition, one can find solutions for both positive and negative $\alpha$ (see Fig.~\ref{positive}).

This fact, in together with the discussion of Higgs fit improvement in the last section, motivated us to look into the case $\alpha<0$, for its possible impact on the phase transition. 
With $\alpha<0$, the mass cubic term $- T m_s^3(\phi, T)$ grows with $\phi$. Therefore, with $S$ itself the degenerate minima or first-order phase transition cannot happen. Namely, with such $S$ itself the equation $V'(\phi, T)=0$ has not solution at non-zero $\phi$, if the condition for building the barrier, $V''(0, T)>0$, is to be fulfilled at the same time.

However, in the environment of degenerate minima created by the other fields, the presence of such $S$ can further enlarge the ratio $v_c/T_c$. This picture can be realized if there are multiple relevant particles during the phase transition. 

As an example, we introduce two color triplet particles $S_1$, $S_2$, with electric charges $2/3$ and $-1/3$ respectively.
We fix the parameters $\alpha_1=0.5$, $m_{s_2}=130\,$GeV and vary the others, $\alpha_2$ and $m_{s_1}$.
Here $S_1$ is designed to mimic the light stop in the MSSM.
For $\alpha_2>0$, the self interactions of $S_{1,2}$ are fixed to be $\kappa_1\approx\kappa_2\approx g_3^2/6$. 
For $\alpha_2<0$, we take into account of the constraint that the zero-temperature potential should be bounded from below at infinity, and modify the value of $\kappa_2$ to satisfy $\kappa_2>\alpha_2^2/\lambda$, which deviates from the MSSM value.
The Higgs self interaction $\lambda=0.13$ is fixed by its mass 125\,GeV, and $\kappa_2$ is bounded from above ($\sqrt{4\pi}$) by perturbativity.
These in turn gives a lower limit $\alpha_2\gtrsim-0.7$.

In the numerical calculation, we have taken into account of the Coleman-Weinberg potential, $V_{\rm CW}$, as well as the higher order logarithmic terms in the finite temperature potential, $\delta V_T$. They are
\begin{eqnarray}
V_{\rm CW}\left(\phi\right) &=& \sum_B n_B  \frac{m_B^4(\phi)}{64\pi^2}\left[ \ln \frac{m_B^2(\phi)}{\Lambda^2} - c \right] + (-1) \sum_F n_F \frac{m_F^4(\phi)}{64\pi^2}\left[ \ln \frac{m_F^2(\phi)}{\Lambda^2} - c \right] \ ,\nonumber \\
\delta V_T (\phi, T)&=& - \sum_B n_B \frac{m_B^4(\phi)}{64\pi^2} \left[ \log \frac{m_B^2(\phi)}{T^2} - c_B \right] +  \sum_F n_F \frac{m_F^4(\phi)}{64\pi^2} \left[ \log \frac{m_F^2(\phi)}{T^2} - c_F \right],
\end{eqnarray}
where $c_B=5.41$, $c_F=2.64$, $c=3/2$ for scalar and fermions, and $5/6$ for gauge bosons. 
The total potential is therefore
\begin{eqnarray}
V_{\rm tot}(\phi, T) = V(\phi, T) + V_{\rm CW}(\phi) + \delta V_{T}(\phi, T) \ .
\end{eqnarray}
We realize the analysis relying on the high temperature expansion at this precision may be subjected to corrections from e.g., two-loop corrections \cite{Cohen:2011ap} and the issue of gauge dependence~\cite{Patel:2011th, Wainwright:2011qy, Wainwright:2012zn}, and leave a more complete analysis to a future study.

\begin{figure}[t!]
\centering
\includegraphics[width=0.8\textwidth]{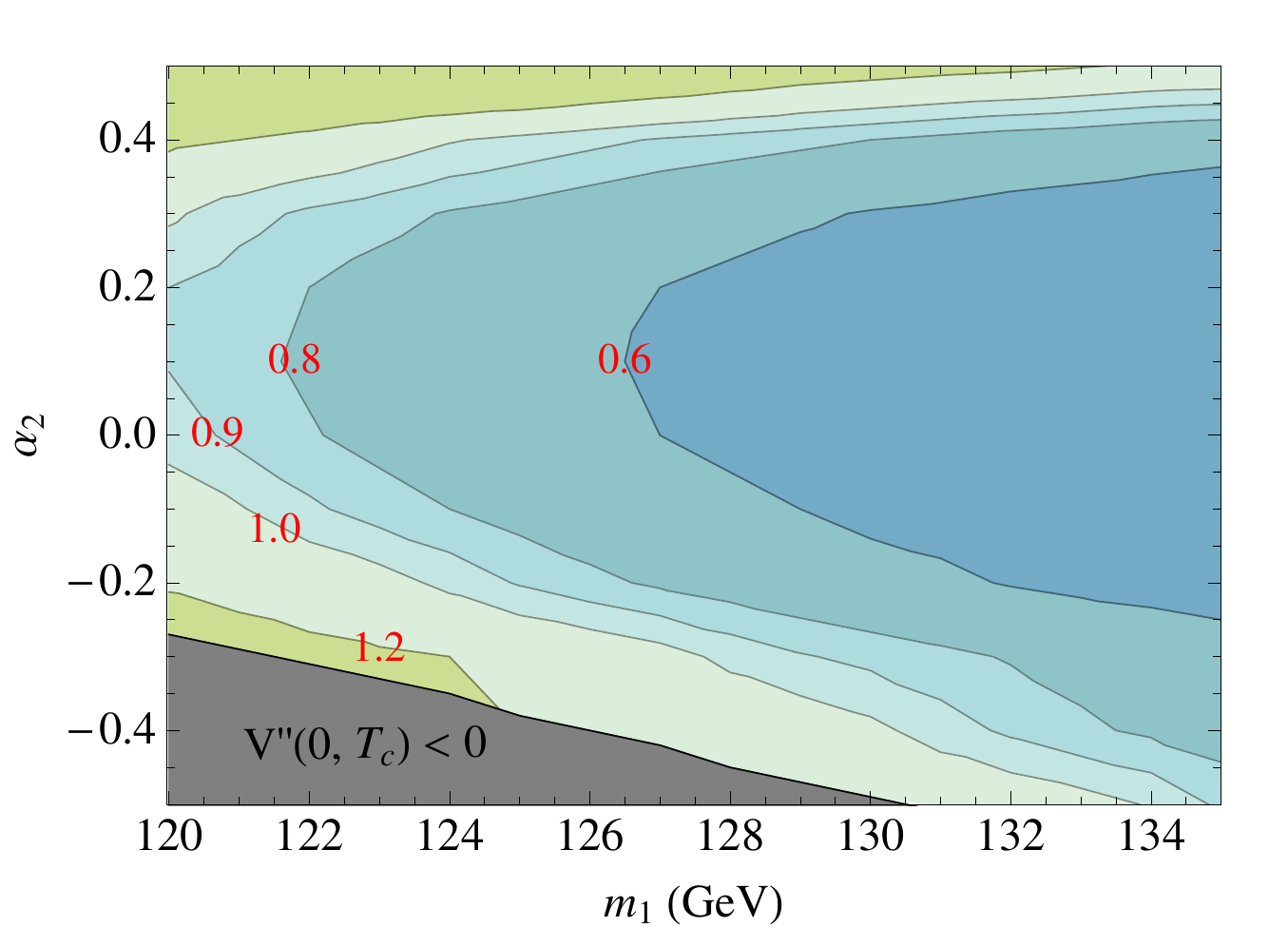}
\caption{The ratio of $v_c/T_c$ in a toy model with two color triplet scalars $S_1$ and $S_2$ coupling to the Higgs field. We fix $\alpha_1=0.5$, $m_{s_2}=130\,$GeV and vary $m_1$, $\alpha_2$. It turns out $v_c/T_c>0.9$ can be achieved for both positive and negative $\alpha_2$, when $m_{s_1}>120\,$GeV. The gray region is excluded because the condition $V''(\phi=0, T_c)>0$ is violated.}\label{toymodel}
\end{figure}

Fig.~\ref{toymodel} shows the contours of the ratio $v_c/T_c$.
When $\alpha_2=0$, we find the strong phase transition condition $v_c/T_c \gtrsim 0.9$ is satisfied at $m_{s_1}\lesssim 120\,$GeV. 
This reproduces the result found in Ref.~\cite{Carena:2008vj}.
On top of this, introducing a positive $\alpha_2$ will further enhance the ratio as expected.

An interesting feature we want to highlight is, in the case $\alpha_2<0$, stronger phase transition is also achievable.
This is due to the positiveness of $F[m_s]$ -- an additional scalar with any sign of $\alpha$ may strengthen the phase transition, as long as it is already made first order.
This enhancement can also be qualitatively understood as follows.
Around the critical temperature, the $S_1$ mass cubic term (with large enough $\alpha_1$) in the thermal potential drags $V(\phi)$ down toward the second minimum.
Meanwhile the $S_2$ cubic term operates in the opposite direction and tends to postpone this to happen, 
until it drops out from the thermal potential (real part). This leads to a larger value for the critical VEV $v_c$. 

In order for the barrier between $\phi=0$ and $\phi=v_c$ minima to be built, we again impose the criterion $V''(\phi=0, T)>0$, 
which also helps to remove the appearance of additional local minimum~\cite{Cline:1996mga}.
This sets another relevant lower bound on $\alpha_2$ (see Fig.~\ref{toymodel}). 
It in turn implies an upper bound on $v_c/T_c$ for negative $\alpha_2$.
There is also an upper bound on the mass of the second scalar, $m_{s_2}<1.8\, T_c$, which is roughly 200\,GeV.
This has to be satisfied in order for the distribution of $S_2$ not to be Boltzmann suppressed in the plasma.

We find for $\alpha_2\approx -0.5$, the mass of $S_1$ can be enlarged up to 135--140\,GeV, while maintaining EWPT strong.
These features are clearly welcome by the Higgs global fit, as discussed in the previous section.

We notice that in Fig.~\ref{toymodel}, the contour of $v_c/T_c$ is shifted upwards and minimal value of phase transition strength for a given $m_1$ is around $\alpha_2 \sim 0.1$. The reason for this asymmetric position of the $v_c/T_c$ contour can be attributed to the impact of $\alpha_2$, which appears in the thermal masses of Higgs boson, Goldstones, and the scalar $S_2$. Their masses can be generally parametrized as
\begin{equation}
m^2 \sim m^2 + \frac{1}{12}(\alpha_2 + \dots)T^2 + \alpha_2 v^2
\end{equation}
like Eq. (\ref{9}) where we have highlighted the $\alpha_2$ contribution via temperature and the electroweak VEV. After daisy resummation, this is the mass that appears in the thermal cubic potential. The $\alpha_2 v^2$ term, if dominate the thermal mass, makes additional contribution to the ratio $v_c/T_c$, while the $\alpha_2 T^2$ part tends to dilute this effect. There is a competition between the two effects, and numerically this leads to the shift/asymmetry observed in Fig.~\ref{toymodel}.

\section{An Application to MSSM and Beyond}

At this point, one is certainly tempted to consider the realization of the above framework in supersymmetric theories, 
especially in view of the tension observed recently between the Higgs fit and the light stop window for 
EWBG~\cite{Cohen:2012zza, Curtin:2012aa, Carena:2012xa, LTWang}.
One would expect additional light colored scalar (e.g., the sbottom, see below) with similar mass to the stop but opposite couplings to the Higgs boson.
Such light sbottom plays double roles, i.e., to improve the fit to Higgs data and to strengthen the EWPT.
Notice that a light sbottom may also contribute to the equilibrium transportation and CP violation sources in EWBG \cite{Chung:2008aya}.

Hereafter, we assume the other components in the Higgs sector, $H$, $A$, $H^\pm$, are much heavier than the lightest scalar $h$. In the decoupling limit, the tree-level couplings of $h$ to weak gauge bosons and fermions are the same as in the SM. 

\subsection{Light sbottom assisted light stop scenario}

We first briefly review the light stop scenario for a strong EWPT. The mass matrix for stops $(\tilde t_L, \tilde t_R)$ is
\begin{eqnarray}
\left[\begin{array}{cc}
m_Q^2 + m_t^2 + D_L^{t} & y_t \phi( A_t \sin \beta - \mu \cos\beta) \\
y_t \phi( A_t \sin \beta - \mu \cos\beta) & m_U^2 + m_t^2 +D_R^t
\end{array} \right] \ , 
\end{eqnarray}
where $m_t(\phi)=y_t \phi \sin\beta$ and for a given fermion $f$, the D-term mass $D^f = (T_f^3 -Q_f s_W^2) \cos (2 \beta) M_Z^2$. In the limit $m_Q\gg m_U$, the light stop mass is
\begin{eqnarray}
m^2_{\tilde t_1} \approx m_U^2 +D_R^t + y_t^2 \sin^2 \beta \left(1-\frac{X_t^2}{m_Q^2}\right) \phi^2 \ \ \ \Rightarrow \ \ \ \alpha_{\tilde t_1} = y_t^2 \sin^2 \beta \left(1-\frac{X_t^2}{m_Q^2}\right) \ ,
\end{eqnarray}
where $X_t=A_t - \mu \cot\beta$. The heavier stop mass is $m_{\tilde t_2} \approx m_Q$, much larger than the electroweak scale.
We take $\tan\beta \gg1$ and $X_t \ll m_Q$ throughout the paper. 
During the electroweak phase transition, $\tilde t_1$ contributes to the mass cubic potential term in Eq.~(\ref{pot}). 
The thermal mass is $m^2_{\tilde t_1} (\phi, T) = m^2_{\tilde t_1} + \Pi_R(T)$.
The light stop scenario for strong electroweak phase transition corresponds to the condition~\cite{Carena:1996wj},
$m_U^2 + \Pi_{\tilde t_R}(T) \approx 0$, or $m_U^2<0$.
In this case, the high temperature mass $m^2_{\tilde t_1} (\phi, T) \approx y_t^2 \sin^2 \beta \phi^2$, and the top Yukawa coupling is large enough to deliver $v_c/T_c\gtrsim 0.9$. An upper bound on the zero-temperature light stop mass is found in~\cite{Carena:2008vj}, which is around 120\,GeV.
As discussed in Sec.~\ref{FIT}, this brings confliction with the current Higgs data.

In order to get out of this dilemma, we apply the framework discussed in the previous sections and bring in additional light colored scalar with $\alpha<0$.
In the MSSM, the only possible candidate is the sbottom. 
Their mass matrix in the basis of $(\tilde b_L, \tilde b_R)$ is
\begin{eqnarray}
\left[\begin{array}{cc}
m_Q^2 + m_b^2 + D_L^b & y_b \phi ( A_b \cos \beta - \mu \sin\beta) \\
y_b \phi ( A_b \cos \beta - \mu \sin\beta) & m_D^2 + m_b^2 +D_R^b
\end{array} \right] \ ,
\end{eqnarray}
where $m_b(\phi)=y_b \phi \cos\beta$. Notice the same $m_Q$ appears, because $\tilde b_L$ and $\tilde t_L$ belong to the same $SU(2)_L$ doublet.
Therefore the heavier sbottom has mass $m_{\tilde b_2}\approx m_Q$ and will be nearly degenerate with the heavier stop.
A crucial difference from the stop case is that, large coupling to the Higgs VEV can be only obtained through large sbottom mixings, 
i.e., $X_b = A_b \cot \beta - \mu$ can be large if the $\mu$ parameter is large. For $m_Q^2\gg m_D^2$, the lightest sbottom state has mass
\begin{eqnarray}
m^2_{\tilde b_1} \approx m_D^2 + D_R^b - \frac{m_b^2 \tan^2\beta}{v^2} \frac{X_b^2}{m_Q^2} \phi^2 \ \ \ \Rightarrow \ \ \ 
\alpha_{\tilde b_1} =- \frac{m_b^2 \tan^2\beta}{v^2} \frac{X_b^2}{m_Q^2} <0 \ .
\end{eqnarray}
For large $\tan\beta$ and $\mu$, negative and sizable $\alpha_{\tilde b_1}$ can be obtained. 
Notice if such sbottom was present, the stop can be made heavier, so the danger of our universe may 
develop into the color breaking vacuum~\cite{Carena:1996wj} is less severe.

\subsection{Constraints in the MSSM}\label{constraints}

Here we show there is a series of constraints in the minimal model that prevent the above light sbottom assisted light stop scenario to work as designed at the beginning of this section.

First, at large field values, the zero-temperature potential is approximately
\beq
V(\phi, \tilde b_1) \approx \frac{\lambda}{4} \phi^4 - \alpha_{\tilde b_1} \phi^2 \tilde b_1^* \tilde b_1 + \kappa_{\tilde b_1} (\tilde b_1^* \tilde b_1)^2 \ .
\eeq
For the potential to be bounded from below, we must require 
\bea\label{LB}
\alpha_{\tilde b_1} > -\sqrt{\kappa_{\tilde b_1}\! \lambda} = - \sqrt{\frac{1}{6}\left( g_3^3 + \frac{1}{3}g_1^2 \right) \cdot \frac{m_h^2}{2v^2}} \approx -0.2 \ .
\eea
Here the self interaction $\kappa_{\tilde b_1} = \kappa_{\tilde b_R} \approx 0.26$ is fixed by the gauge couplings from the $\tilde b_R$ D-terms. There is no room to enlarge it if MSSM is taken to be a complete theory.

The second but more severe tension arises from a combined consideration of Higgs mass and vacuum stability.
In the MSSM, radiative corrections are crucial to lift the tree-level Higgs mass. Neglecting the stop mixing effect, we have~\cite{Djouadi:2005gj}
\bea
(m_h^2)_{\rm 1-loop} \approx M_Z^2 \cos^2 2 \beta + \frac{3 G_F m_t^4}{\sqrt2 \pi^2} \log \frac{m_{\tilde t_1} m_{\tilde t_2}}{m_t^2} \ .
\eea
In the light stop scenario $m_{\tilde t_1}<m_t$, in order to achieve $m_h=125\,$GeV, the heavier stop must satisfy $m_{\tilde t_2}\approx m_Q\gtrsim 5\,$TeV.
Since the two loop effects tend to decrease the Higgs mass, we need an even larger $m_Q > 10\,$TeV. \ \
We also consider the vacuum stability~\cite{Reece:2012gi, Kitahara:2012pb, Babu:2002uu, Sato:2012bf, Delgado:2012rk} of the heavy $\tilde b_L$ field. 
Taking $\tilde b_L$ as a background field, there is a negative finite loop correction to its quartic interaction
\bea\label{COR}
\delta \kappa_{\tilde b_L} &\approx& - \frac{1}{32\pi^2} \frac{1}{6} \left(\frac{m_b \tan \beta}{v} \frac{X_b}{\tilde m}\right)^4 
= - \frac{1}{192\pi^2} |\alpha_{\tilde b_1}|^2 \left(\frac{m_Q}{\tilde m} \right)^4 \ ,
\eea
where $\tilde m$ is the mass scale of the particles running in the box diagram, 
$h$ and $\tilde b_R$, both of which are lighter than $\sim 200\,$GeV for our interest. 
In order not to develop a color breaking vacuum of $\tilde b_L$, this correction should be smaller than the
tree-level coupling $\kappa_{\tilde b_L}^{\rm tree} = g_3^2/6 + g_2^2/4 + g_1^2/72\approx0.36$.
For $\alpha_{\tilde b_1} = -0.2$, this implies an upper bound, $m_Q\lesssim 2\,$TeV.

Interchanging the role of left and right-handed sfermion masses does not work either. In this case, the light stop and sbottom will share the same bare mass $m_Q^2$. 
The phase transition requires $m_Q^2\lesssim0$ from the stop side, while the sbottom (with $\alpha_{\tilde b_1}<0$) satisfying the LEP limit requires $m_Q^2\gtrsim 100\,$GeV. Moreover, this case is also more constrained by the electroweak precision $T$ parameter~\cite{Espinosa:1993yi}.

To summarize, the above discussions exclude $\alpha_{\tilde b_1}$ from being sizable, and reinforce the status that MSSM electroweak baryogenesis becomes disfavored after the Higgs discovery.

\subsection{Possible solutions by going beyond}

The above constraints could be evaded if the MSSM is regarded as an effective theory, plus the remnant effects from higher scale physics.
Possible solutions to the vacuum stability problem include enhancing the self interactions of Higgs boson and $\tilde b_L$. 
If the $\tilde b_R$ self interaction is also enlarged, the lower bound in Eq.~(\ref{LB}) can be further relaxed. 

One may argue that the light stop window is less unique if one is allowed to go beyond MSSM. However, if the sbottom assisted light stop scenario were to work, a clear prediction would be very light third generation sfermions, which are testable even at the current stage of LHC, as emphasized in the next subsection.

The extra contributions to the Higgs mass allow the second stop mass $m_Q$ to be substantially smaller, so the dangerous correction in Eq.~(\ref{COR}) can be made safely small. 
The enhancement in the Higgs boson self-interaction widely exists in the extensions to the MSSM, from the F-term~\cite{Delgado:2012rk}
such as the NMSSM and , or
from a non-decoupled D-term~\cite{Batra:2003nj, Maloney:2004rc, Zhang:2008jm}~\cite{Endo:2011gy, Cheung:2012zq, An:2012vp} (see also~\cite{Babu:1987kp}). 
Here, we show an example following the discussion of supersymmetric left-right model in~\cite{Zhang:2008jm}. 
The gauge symmetry is $SU(2)_{L}\times SU(2)_R\times U(1)_{B-L}$ which breaks to SM gauge symmetry at a scale $v_R\gtrsim$ TeV. 
In order to match to the SM D-terms in the supersymmetric decoupling limit, the tree-level exchange of the uneaten scalar component (real part) in the would-be Goldstone superfield is found to be crucial. Soft SUSY breaking mass terms violate its degeneracy with the heavy gauge bosons and gauginos.
The correction to the Higgs self interaction is 
\bea
2 \delta \lambda_h = \frac{g_2^2 g_{BL}^2 + g_2^2 \frac{m^2}{v_R^2}}{g_2^2 + g_{BL}^2 + \frac{m^2}{v_R^2}} - g_1^2 \ ,
\eea
where $g_{BL}^2 = g_2^2 g_1^2/(g_2^2 - g_1^2)$, and $m$ here is the soft SUSY breaking mass in the $SU(2)_R\times U(1)_{B-L}$ breaking Higgs sector.
For $m^2/v_R^2=2$ and $m_{\tilde t_1}=130\,$GeV, $m_{\tilde t_2}\approx m_Q \approx 2\,$TeV, we find 125\,GeV Higgs mass can be easily accommodated.
This solves the second constraint discussed in Sec.~\ref{constraints}.
 
The self-interaction of sbottoms may be enhanced through the F-term 
by coupling them to new vector-like states. 
As an illustration, we introduce a pair of exotic states $X$, $\bar X$ and the corresponding superpotential $W = M X \bar X + \lambda' X b^c b^c$.
After integrating out $X, \bar X$, there could be a correction to the $\tilde b_R$ self coupling, in the presence of soft SUSY breaking mass $m$ of $\bar X$,
$\delta \kappa_{\tilde b_R} = |\lambda'|^2 m^2/(M^2+m^2)$. If $M\sim m$ and $\lambda'$ is large, this helps to relax the first constraint above.

\subsection{Light sbottom and stop at LHC}

The main testable prediction from the above picture, is the existence of a light sbottom (below 200\,GeV), and a relatively heavier light stop (can be as heavy as 140\,GeV). The light sbottom and stop can be directly and copiously produced at hadron colliders. This opens up further motivations for the urgency of their direct searches at the LHC in the near future.

Direct search has been regarded as a crucial test of the light stop scenario for EWBG. 
The current limits are summarized and discussed in~\cite{Carena:2012xa, ATLASstop}, which vary depending on the decay channels.
For the stop mass below $<150\,$GeV, it mainly decays into $\tilde t_1 \to b W^+ \tilde \chi_1^0$, via an off-shell chargino or top quark,
where the dominant background from SM $t\bar t$ production.
We observe there is no limit from Tevatron or LHC for a light stop whose mass is near to that of the lightest neutralino~\cite{ATLAStalk}.
In this case, the lower limit comes from LEP, which is 96\,GeV. 
The bound could get stronger when the mass difference between stop and neutralino is close enough to the co-annihilation regime and $\tilde t_1 \to c\tilde \chi_1^0$ takes over. It seems challenging to exclude the EWBG scenario by searching for the stop alone.

This gives more priority to look into the light sbottom direct searches, which is another prediction from the framework set up in this paper.
A light sbottom could decay into a bottom quark and the lightest neutralino, $\tilde b_1\to b\tilde \chi_1^0$. The current lower bound for this channel using events with two hard b-jets and missing energy can be as large as 400\,GeV, for a very light neutralino~\cite{Aad:2011cw}. However, this bound gets much weaker, down to $\sim 100\,$GeV when the mass difference between sbottom and neutralino are close.
The monojet events are expected to play a complementary role~\cite{Aaltonen:2012jb, Chatrchyan:2012me}, and have
the potential to cover the whole light sbottom region after the 8\,TeV run of LHC~\cite{Alvarez:2012wf}. 
The sbottom may also be longer lived than the collider time scale. In this case, it can also be tightly constrained by searching for events with displaced vertices or even stopped particles inside the detectors~\cite{Fairbairn:2006gg, :2012yg}.

\section{Conclusions}

The discovery of a 125\,GeV Higgs boson at LHC has led us into an era of precise measurement of Higgs couplings.
This will finally give us deeper understanding of the EWSB and possible connections to new physics from the Higgs portal.
We confront the new electroweak states that can trigger strong first order EWPT to the global fit of the current Higgs data.
A close correlation exists between the two phenomena if such new states carry color and/or electric charge.
It has been pointed out that a single scalar responsible for strong EWPT, such as a light stop in the MSSM, is in tension with the current Higgs fit.

In this work, we extend this minimal picture and study the possibility of having multiple new scalars present at the electroweak scale.
We set up a general framework, in which a first scalar $S_1$ with positive Higgs portal coupling triggers the first-order phase transition, 
and a second scalar $S_2$ with similar mass but opposite Higgs coupling not only improves the Higgs fit, 
but also further enhances the strength of phase transition. This enables us to substantially improve the compatibility between the Higgs fit and EWPT, which we notice is essentially independent of how the first order EWPT is triggered. Accordingly, we expect the mechanism proposed here extremely intriguing, and widely applicable in almost all beyond SM models with a first order EWPT.

We have discussed possible realization of this framework in SUSY theories, including the MSSM and beyond.
We realize the constraints in the MSSM from the color breaking vacua, and find they could be evaded by extending the minimal model.
We show two examples where the Higgs boson and the squark self interactions can be enhanced through non-decoupled D-term and/or F-term.
The prediction of both a light stop (around 140\,GeV) and a light sbottom (below 200\,GeV) is highlighted, and 
which can be directly tested at the LHC in the near future.

\section*{Acknowledgements}
We thank K.S. Babu, Lotfi Boubekeur, Rabi~Mohapatra, Miha Nemev\v{s}ek, Goran Senjanovi\'{c} for illuminating discussions and comments on the manuscript. 
We acknowledge the code for the Higgs global fit from Da Liu.


\begin{thebibliography}{99}
\bibitem{comment1}
Fermions with large yukawa couplings to Higgs can also trigger a strong first order EWPT if we consider their entropy production when they are decoupled. See~\cite{entropy}. Nevertheless, this mechanism requires some fine tuning of the Higgs potential and is not generic so we do not consider this possibility here.
\end{thebibliography}
\end{document}